\documentclass[aps,pre,superscriptaddress,amsmath,amssymb,twocolumn]{revtex4}
    \usepackage{graphicx}
    \usepackage{hyperref}
    \usepackage{color}
    \definecolor{darkblue}{rgb}{0.0,0.0,0.5}
    \hypersetup{colorlinks,breaklinks,
                linkcolor=darkblue,urlcolor=darkblue,
               anchorcolor=darkblue,citecolor=darkblue}
    
    \date{\today}
    
    \allowdisplaybreaks
    
\begin{document}
    
    \title{Dimer Covering and Percolation Frustration}
    
    \author{Amir Haji-Akbari}
    \email{hajakbar@princeton.edu}
    \affiliation{Department of Chemical and Biological Engineering, Princeton University, Princeton NJ 08540}
    
    \author{Nasim Haji-Akbari}
    \email{nasimh@umich.edu}
\affiliation{Department of Chemical Engineering, University of Michigan, Ann Arbor, MI 48109}
    
    \author{Robert M Ziff}
    \email{rziff@umich.edu}
    \affiliation{Department of Chemical Engineering, University of Michigan, Ann Arbor, MI 48109}
    
    \date{\today}
    
    \begin{abstract}
    Covering a graph or a lattice with non-overlapping dimers is a problem that has received considerable interest in areas such as discrete mathematics, statistical physics, chemistry and materials science. Yet, the problem of percolation on dimer-covered lattices has received little attention. In particular, percolation on lattices that are fully covered by non-overlapping dimers has not evidently been considered. Here, we propose a novel procedure for generating random dimer coverings of a given lattice. We then compute the bond percolation threshold on random and ordered coverings of the square and the triangular lattice, on the remaining bonds connecting the dimers. We obtain $p_c=0.367713(2)$ and $p_c=0.235340(1)$ for random coverings of the square and the triangular lattice, respectively. We  observe that the percolation frustration induced as a result of dimer covering is larger in the low-coordination-number square lattice.  There is also no relationship between the existence of long-range order in a covering of the square lattice, and its percolation threshold. In particular, an ordered covering of the square lattice, denoted by shifted covering in this work, has an unusually low percolation threshold, and is topologically identical to the triangular lattice. This is in contrast to the other ordered dimer coverings considered in this work, which have higher percolation thresholds than the random covering. In the case of the triangular lattice, the percolation thresholds of the ordered and random coverings are very close, suggesting the lack of sensitivity of the percolation threshold to microscopic details of the covering in highly-coordinated networks. 
    \end{abstract}
    
    \maketitle

    \section{Introduction\label{section:intro}}
    
    Percolation deals with the emergence of long-range connectivity in random systems, and is widely utilized in studying phenomena such as conductivity~\cite{TsuAPL1982}, magnetism~\cite{ElmersPRL1994}, porous fluid flow~\cite{SahimiRevModPhys1993}, permeation~\cite{SchmittmannJMathChem2008}, jamming and glass transition~\cite{ToninelliPRL2006}, gelation~\cite{StaufferJChemSocFarTrans1976} and disease propagation~\cite{MeyersBullAmerMathSoc2007}. To that end, lattices and ordered networks are excellent model systems for studying percolation, with randomness introduced through using a known parameterized probability distribution for occupying the bonds and/or sites of the corresponding network. The manifold in the parameter space of the probability distribution that corresponds to the emergence of long-range connectivity in the network is typically referred to as the \emph{percolation threshold}, and has been computed analytically or numerically for a wide range of lattices and networks in different dimensions~\cite{EssamJMathPhys1964, DjordjevicJPhysAMath1982, AdlerPRB1990,  BabalievskiPhysicaA1995, ZiffJPhysA1997, vanDerMarckIntJMPC1998, LorenzZiffPRE1998,  SudingZiffPRE1999, ZiffPhysicaA1999,  HuangPRE1999, NewmanZiffPRE2001, BertinAdcApplProv2002,  GalamPRE2005, ZiffPRE2006,  ScullardPRE2006, ParviainenJPhysAMath2007, BlotePRE2008, HajiAkbariPRE2009, BeckerPRE2009, ZiffJStatMech2010, SedlockJPhysAMath2012, SmallJStatMechThExp2013, MelchertPRE2013, JacobsenJPhysA2014, SuszczyskiJStatMech2014, MalarzPRE2015}.
    
    A particular class of lattices with unknown percolation characteristics are the dimer-covered lattices. In discrete mathematics, a \emph{dimer covering} of a graph is a selection of edges of the graph so that each vertex is connected to exactly one such edge. Dimer coverings are of immense interest to fields such as discrete mathematics, statistical physics, chemistry and materials science. The problem of enumerating the number of distinct dimer coverings of a graph has received considerable interest in the literature~\cite{KasteleynPhysica1961}, especially in the asymptotic limit of infinitely large lattices~\cite{HammersleyPCPS1968}. A conceptually related problem in statistical mechanics is to determine the partition functions of on-lattice systems that can be mapped onto the dimer covering problem.  For instance, there is a correspondence between the fully frustrated Ising model and the dimer covering of its dual lattice~\cite{HenleyJStatPhys1997}. Dimer coverings of finite honeycomb motifs are known as Kekul\'{e} structures in chemistry, and are related to the number of ways that double bonds can be distributed on a polycyclic aromatic carbon backbone~\cite{SchmalzTCA1986}. In materials science, dimer coverings arise in phases that assemble by dimers of a particular building block. Such phases that are sometimes referred to as \emph{degenerate (quasi)crystals}~\cite{WojciechowskiPRL1991, HajiAkbariDQC2011} are aperiodic and are the dimer coverings of the crystals~\cite{AlderWainwright1957} or the quasicrystals~\cite{HajiAkbariEtAl2009,HajiAkbaricondmat2011} assembled by the corresponding monomers. These degenerate phases are stabilized by the pairing entropy, i.e.,~the entropy associated with  distinct ways of covering a network with dimers. They are of particular interest because of their superior mechanical properties in comparison to their non-degenerate monomer-based counterparts~\cite{CohenPRL2008, GerbodeCohenPRL2010}. Degenerate phases of $n$-mers have also been observed in polymeric systems~\cite{KarayiannisLasoSoftMatter2010, KarayiannisLasoPRL2009}, which can be conceptually considered as an extension of the dimer covering problem to situations where each vertex of the graph is  part of a tree with $n$ vertices. 
    
    Despite the large body of work on enumerating the combinatorics of dimer-covered graphs and  the partition functions of the associated model systems, very little is known about  percolation on dimer-covered lattices. All existing studies have been carried out on lattices partially covered with dimers/$k$-mers by random sequential addition processes, and considering site percolation~\cite{BecklehimerPhysicaA1992, PommiersPRB1994, GaoPhysicaA1998, KondratPRE2001, NietoPhysicaA2003, NietoEPJB2003,  CherkasovaEPJB2010, LaptevPRE2012, VrhovacPRE2012}, or on systems such as interacting dimers~\cite{LjPhysiaA2014}, and no numerical estimate of the percolation threshold has been obtained for fully dimer-covered lattices. Apart from the percolation threshold itself, a particular question of interest is the relationship between the percolation on a dimer-covered lattice, and the percolation on the associated uncovered lattice. As will be explained further in Section~\ref{section:statement}, the constraint of dimers being non-overlapping will introduce some frustration in the percolation of the original lattice. In other words, the total number of bonds that need to be occupied for the lattice to percolate will in general be higher in the dimer-covered lattice. However, the extent of such frustration is not known \emph{a priori} and can be different from lattice to lattice, and from covering to covering.

    Another point of interest is the inherent aperiodicity of  random dimer coverings of a lattice. In other words, the topologically equivalent network of a random covering (see Section~\ref{section:statement} for details) is a random network itself. Therefore, the dimer covering process can be viewed as a novel way of generating random networks. Particularly, the construction of such networks is purely topological and does not require the definition of a norm. Contrast this to other known examples of random networks that need a notion of distance in their definition. Examples include the relative-neighborhood graph~\cite{MelchertPRE2013}, Gabriel graph~\cite{BertinAdcApplProv2002}, Voronoi tessellation and Delaunay triangulation~\cite{HuangPRE1999, BeckerPRE2009}. Comparing the percolation characteristics of these covering-based networks with the norm-based random networks is also of interest. 
    
    In this work, we study bond percolation on dimer coverings of the square and the triangular lattices using the  Newman-Ziff algorithm~\cite{NewmanZiffPRE2001}. We obtain random coverings of square and triangular lattice using a novel shuffling algorithm that is conceptually similar to an algorithms utilized for creating proton-disordered ice configurations~\cite{StillingerJCP1972}. We assure the randomness of the generated lattices by computing appropriate correlation functions.   We also compute the percolation threshold for several ordered dimer coverings of the square and the triangular lattice for comparison. We refine our estimates from the Newman-Ziff algorithm using the epidemic growth method of Lorenz and Ziff~\cite{LorenzZiffPRE1998}.  We found thresholds to fairly high precision in order to distinguish the values for systems where $p_c$ agreed to three or four digits.
    
    This paper is organized as follows. In Section~\ref{section:statement}, the problem of bond percolation on dimer coverings of a lattice is outlined. The computational methods utilized for generating random coverings, and for measuring percolation thresholds are discussed in Section~\ref{section:methods}. Section~\ref{section:results} is reserved for results and discussions, and concluding remarks are presented in Section~\ref{section:conclusion}.
    
    
    \section{Problem Statement\label{section:statement}}
    Studying percolation on dimer-covered lattices is a two-step process. Consider a lattice $\mathcal{L}$ that is comprised of $N_s$ sites and $N_b$ bonds with $N_s$ being an even integer. The first step involves creating a dimer covering of $\mathcal{L}$, denoted by $\mathcal{L}_d$, in which $(1/2)N_s$ of those $N_b$ bonds are  occupied \emph{permanently} so that each site is connected to a single permanently occupied bond only. The second step involves defining the bond percolation problem on the arising network, $\mathcal{L}_d$, by \emph{stochastically} occupying the remaining $N_b-(1/2)N_s$ bonds with probability $p$, with $p_c$, the bond percolation threshold of $\mathcal{L}_d$ defined as~\cite{ScoppolaJStatPhys2004}:
    \begin{eqnarray}
    p_c &=& \sup\{p:\phi(p)=0\}\label{eq:pc}
    \end{eqnarray}
    Here $\phi(p)$ is the probability of observing an infinite cluster in $\mathcal{L}_d$. For a particular finite covering, the notion of an infinite cluster can be defined on a superlattice obtained from periodically replicating such a covering in all directions. This is the usual way of defining the percolation threshold as the occupation probability at which a percolating cluster emerges for the first time. 
   Here, each cluster is comprised of sites that are connected to one another through a combination of permanently occupied dimer bonds, and stochastically occupied regular bonds, and the definition of $p_c$ is based on the bonds that are occupied stochastically. In other words, the percolation problem defined in (\ref{eq:pc}) is for a new network in which each pair of sites connected through  permanently occupied bonds is united into a single site. Throughout this work, we will refer to this new network as the \emph{topologically equivalent network} associated with the covering. In order to establish the relationship between percolation on the dimer-covered lattice, and on the original uncovered lattice, we define an equivalent percolation threshold as:
    \begin{eqnarray}
    p_{c,\text{eq}} &=& p_c+\frac{N_s}{2N_b}
    \end{eqnarray}
    which is the percolation threshold of $\mathcal{L}$ \emph{conditioned} that the permanently occupied dimer bonds in $\mathcal{L}_d$ are among the bonds that are chosen to be occupied in $\mathcal{L}$. In other words, $p_{c,\text{eq}}$ is the total number of bonds (permanently occupied dimer bonds, and stochastically occupied regular bonds) occupied when $\mathcal{L}_d$ percolates, with $N_s/2N_b=1/4$ and $1/6$ for the square and the triangular lattice, respectively. Note that $\tilde{\phi}_{\text{con},d}(p)\le\tilde{\phi}(p)$ with $\tilde{\phi}_{\text{con},d}(\cdot)$ and $\tilde{\phi}(\cdot)$ the probability of observing an infinite cluster in $\mathcal{L}$ in the case of constrained and unconstrained percolation, respectively. Therefore, $p_{c,\text{eq}}\ge\tilde{p}_c$ with $\tilde{p}_c$ the unconstrained bond percolation threshold of $\mathcal{L}$. The ratio of $\kappa=p_{c,\text{eq}}/\tilde{p}_c$ is a measure of percolation frustration due to the particular covering of $\mathcal{L}$ with non-overlapping dimers. This is a quantity of interest to us and will be computed and discussed in Section~\ref{section:results}. 
     
     It is necessary to emphasize that for any lattice $\mathcal{L}$, a large number of distinct coverings are possible, and for any such covering, $p_c$ can be obtained from Eq.~(\ref{eq:pc}). We define the random covering percolation threshold  as the ensemble average of $p_c$ for all the coverings that can be obtained from $\mathcal{L}$.
     
        \begin{figure*}
    	\centering
		\includegraphics[width=.9\textwidth]{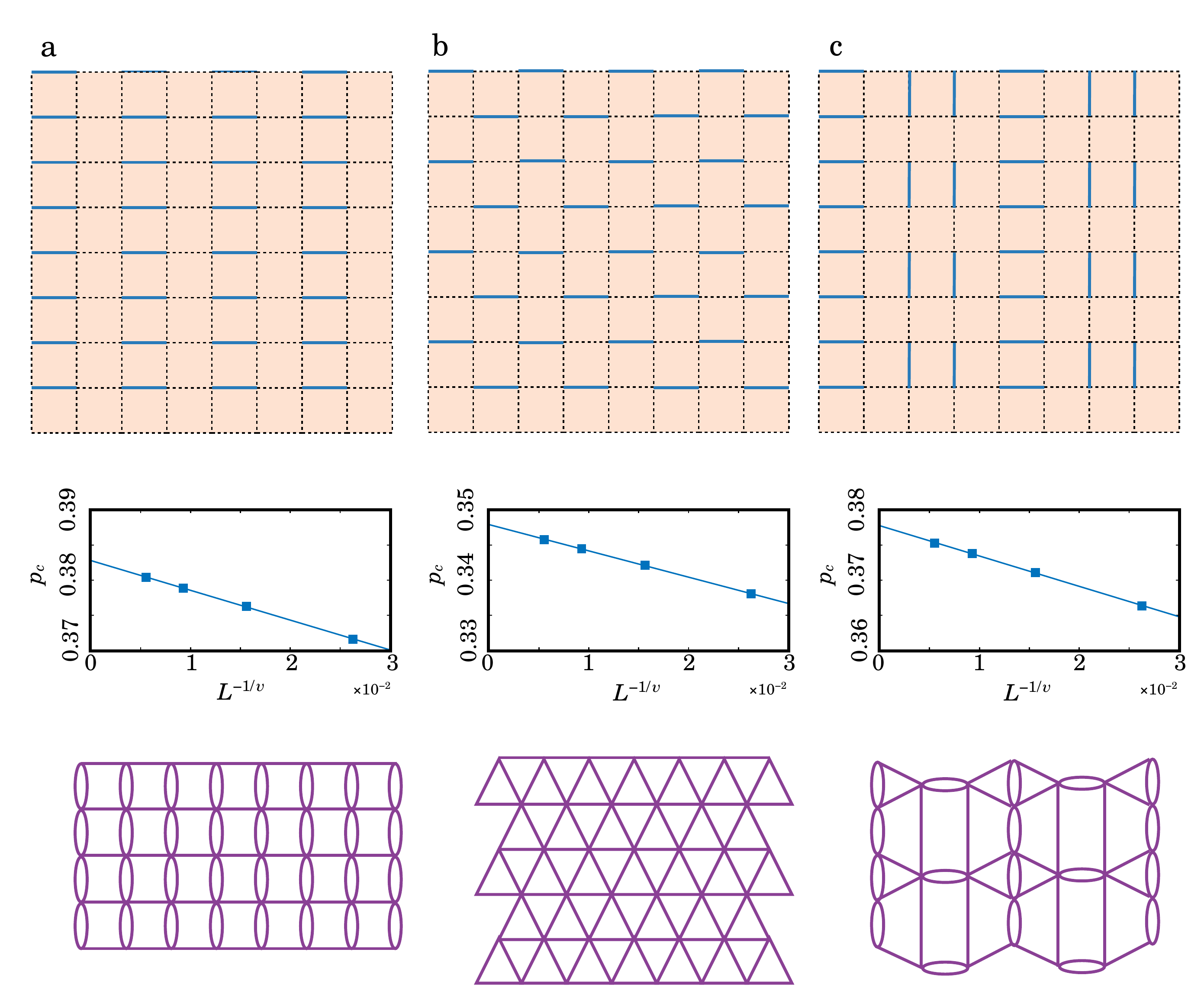}
		\caption{\label{fig:ordered:sq} Bond percolation in (a) parallel, (b) shifted and (c) staggered ordered coverings of the square lattice. The topologically equivalent lattices obtained by uniting the dimer sites are depicted in the bottom. Extrapolations to $L\rightarrow\infty$ have obtained using the linear fit of $p_c$ vs.~$1/L^{1/\nu}$. }
    \end{figure*}
    
    \begin{figure*}
    	\centering
		\includegraphics[width=.7\textwidth]{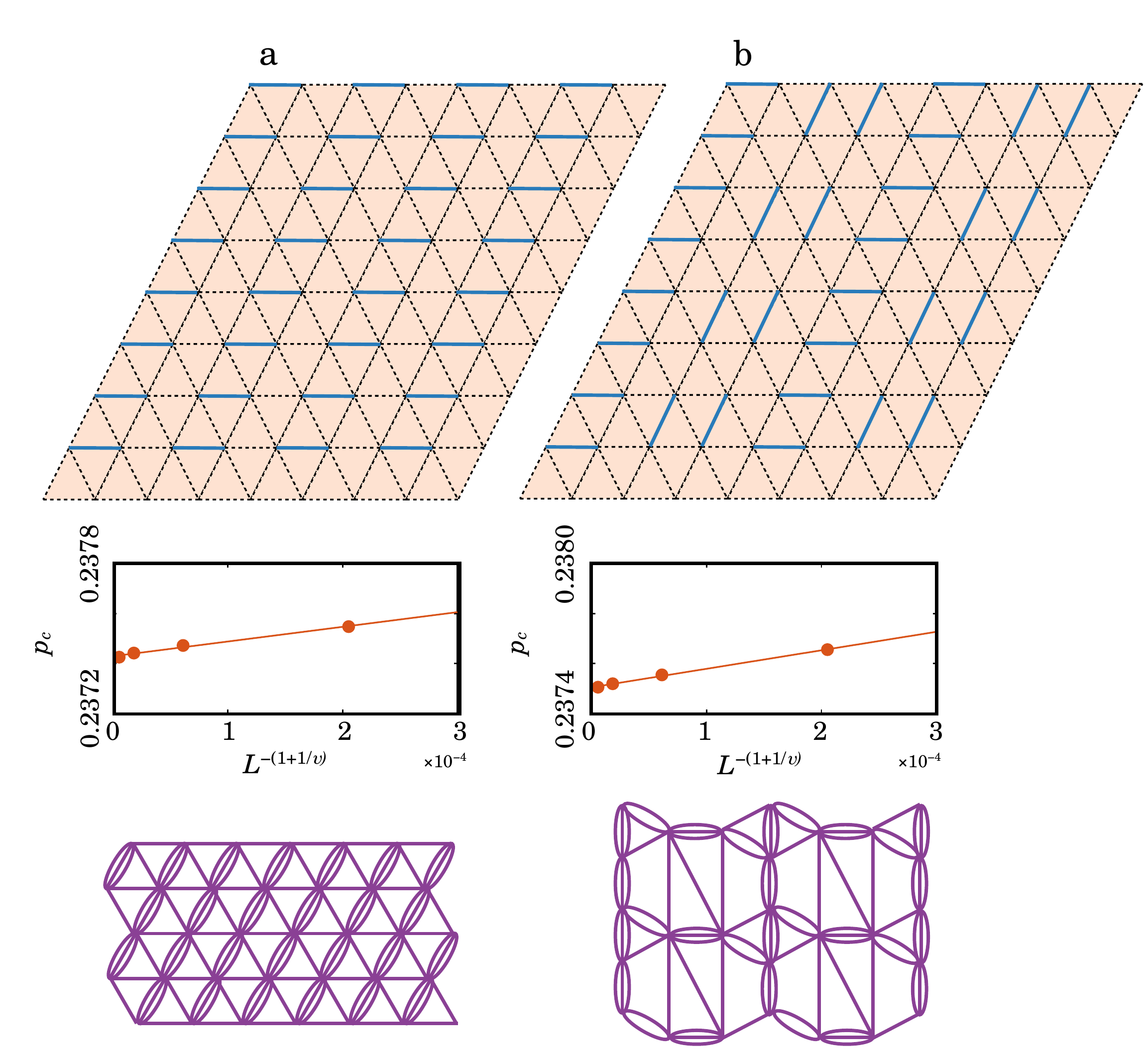}
		\caption{\label{fig:ordered:tr}Bond percolation in (a) parallel and (b) staggered ordered coverings of the triangular lattice. The topologically equivalent lattices obtained by uniting the dimer sites are depicted in the bottom. Extrapolations to $L\rightarrow\infty$ have obtained using the linear fit of $p_c$ vs.~$1/L^{1+1/\nu}$. }
    \end{figure*}
    
        \begin{figure}
    \centering
    \includegraphics[width=.2\textwidth]{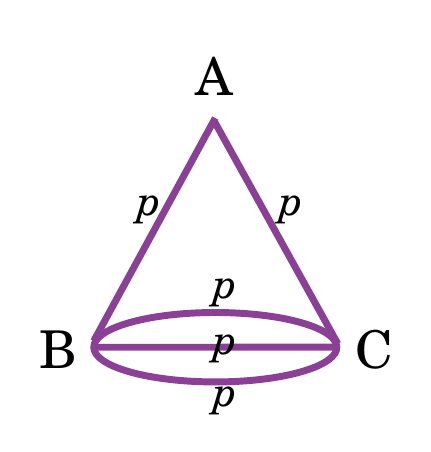}
    \caption{\label{fig:tr_unit}The unit super-triangle of the parallel triangular covering of Fig.~\ref{fig:ordered:tr}a.}
    \end{figure}


   	\section{Methods\label{section:methods}}

    \begin{figure*}
    \centering
    \includegraphics[width=.78\textwidth]{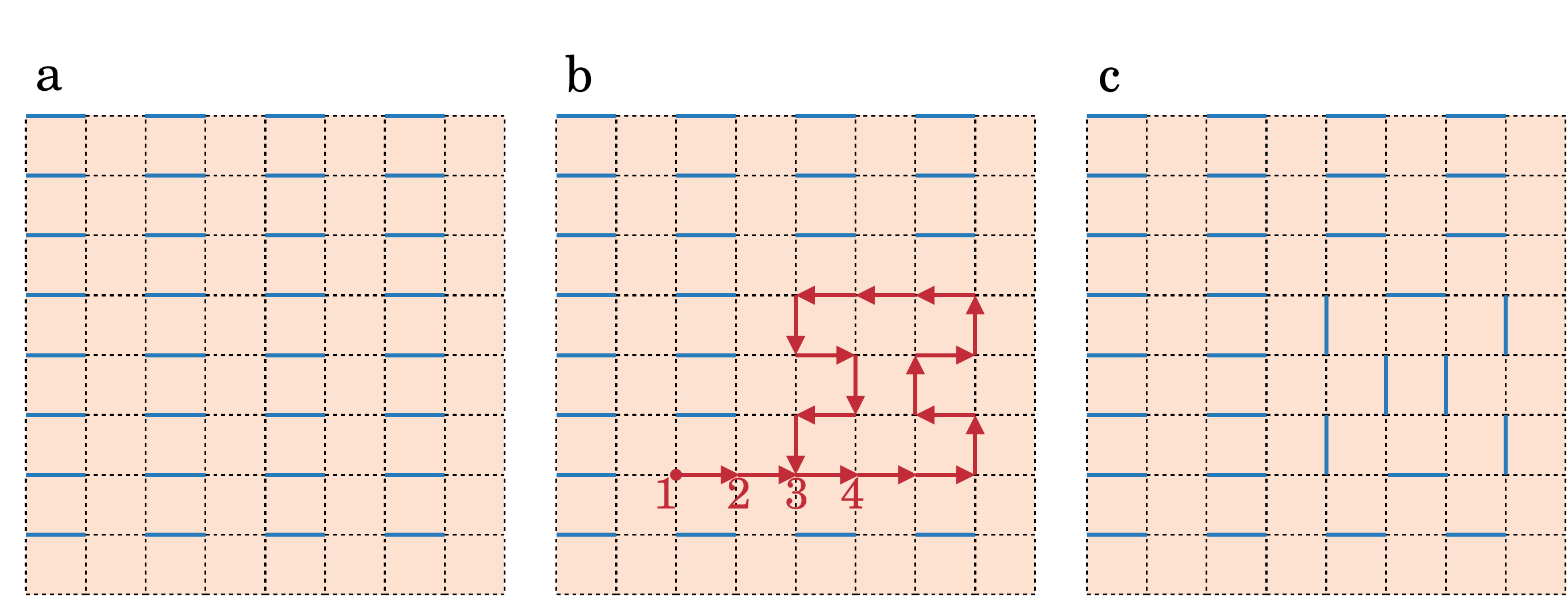}
    \caption{\label{fig:randomize} A representative shuffling move of the algorithm used for generating random coverings of a lattice. Starting with an arbitrary initial covering, such as the parallel covering of (a), a site, such as point 1 in (b), is chosen randomly, and a closed path originating from 1 is constructed by carrying out a random walk. (c) The new covering is obtained by flipping the identity of permanently occupied bonds along the enclosed portion of the path (i.e.,~the portion starting and ending at 3). 
}
    \end{figure*}

	We compute $p_c$ and $p_{c,\text{eq}}$ for several  coverings of the square and the triangular lattice, and we consider both ordered and random coverings. The particulars of the ordered coverings, as well as the procedure used for generating random coverings are thoroughly discussed below. For three of the ordered coverings, $p_c$ can be computed exactly through the triangle-triangle transformation~\cite{ZiffPRE2006,ScullardPRE2006}, or through mapping  onto the problem of inhomogeneous bond percolation on the square lattice. Even for these lattices, we perform the numerical estimation of $p_c$ as a means of validating our computer program. In defining connectivity and covering, we apply periodic boundary conditions in both dimensions. In order to assess $p_c$ in the limit of $N_b\rightarrow\infty$, four lattice sizes are considered: $128\times128$, $256\times256$, $512\times512$ and $1024\times1024$. All random numbers are generated using the four-offset shift-register random number generator, $R(471,1586,6988,9689)$, described in Ref.~\cite{ZiffCompuPhys1998}.

    \subsection{Ordered Dimer-covered Lattices}
    
    For the square lattice, three different ordered coverings are considered. In the \emph{parallel} covering (Fig.~\ref{fig:ordered:sq}a), individual dimer-covered rows are stacked on top of one another. The closely-related \emph{shifted} covering (Fig.~\ref{fig:ordered:sq}b) is also comprised of stacks of dimer-covered rows. However, the positions of dimers are shifted by one lattice site, every other row.  The third covering is the \emph{staggered} covering (Fig.~\ref{fig:ordered:sq}c) and is comprised of horizontal and vertical stripes of the parallel covering.  For the triangular lattice, the same types of coverings are considered. However, here the parallel and shifted coverings correspond to the same lattice (Fig.~\ref{fig:ordered:tr}). We will therefore refer to these two equivalent coverings as the parallel covering from this point onward.

Among these five ordered coverings, the percolation thresholds of three can be computed exactly. As can be noted in the bottom of Fig.~\ref{fig:ordered:sq}b, the shifted covering of the square lattice is topologically identical to the triangular lattice, with its percolation threshold of $p_c=2\sin\frac{\pi}{18}\approx0.347296\cdots$ known analytically~\cite{EssamJMathPhys1964}. The parallel covering of the triangular lattice is also topologically identical to a lattice that can be partitioned into super-triangles. Fig.~\ref{fig:tr_unit} depicts one such super-triangle. Therefore, its percolation threshold can  be computed from the triangle-triangle transformation~\cite{ZiffPRE2006,ScullardPRE2006}  i.e.,~by solving the equation $P(ABC)=P(\overline{ABC})$. Here, $P(ABC)$ is the probability that $A$, $B$ and $C$ in Fig.~\ref{fig:tr_unit} are all connected, while $P(\overline{ABC})$ is the probability that none of $A, B$ and $C$ are connected to one another. For the super-triangle depicted in Fig.~\ref{fig:tr_unit}, $P(ABC)=P(\overline{ABC})$ yields the following algebraic equation:
\begin{eqnarray}
7p^2q^3+9p^3q^2+5p^4q+p^5-q^5=0\label{eq:parallel:tr}
\end{eqnarray}
with $q=1-p$. By solving (\ref{eq:parallel:tr}) numerically, we obtain a percolation threshold of $p_c\approx0.237418\cdots$. This also follows from the anisotropic triangular lattice threshold  splitting one bond into three~\cite{EssamJMathPhys1964}.

The third covering with a $p_c$ known exactly is the parallel covering of the square lattice, which is topologically identical to a square lattice with double bonds in one direction (Fig.~\ref{fig:ordered:sq}a). Bond percolation on such a lattice can be obtained by mapping it onto the problem of inhomogeneous percolation on the square lattice. Suppose a square lattice in which the horizontal and vertical bonds have different occupation probabilities $p_1$ and $p_2$. The percolation manifold of such a lattice is given by $p_1+p_2=1$. For the lattice of Fig.~\ref{fig:ordered:sq}a, let $p_1=p$ and $p_2=1-(1-p)^2=2p-p^2$. The percolation threshold is therefore the root of $p^2-3p+1=0$ which is $p=(3-\sqrt5)/2\approx0.381966\cdots$.

    \subsection{Generation of Random Dimer-covered Lattices\label{section:shuffling}}
    
    \begin{figure*}
    \centering
    \includegraphics[width=.7\textwidth]{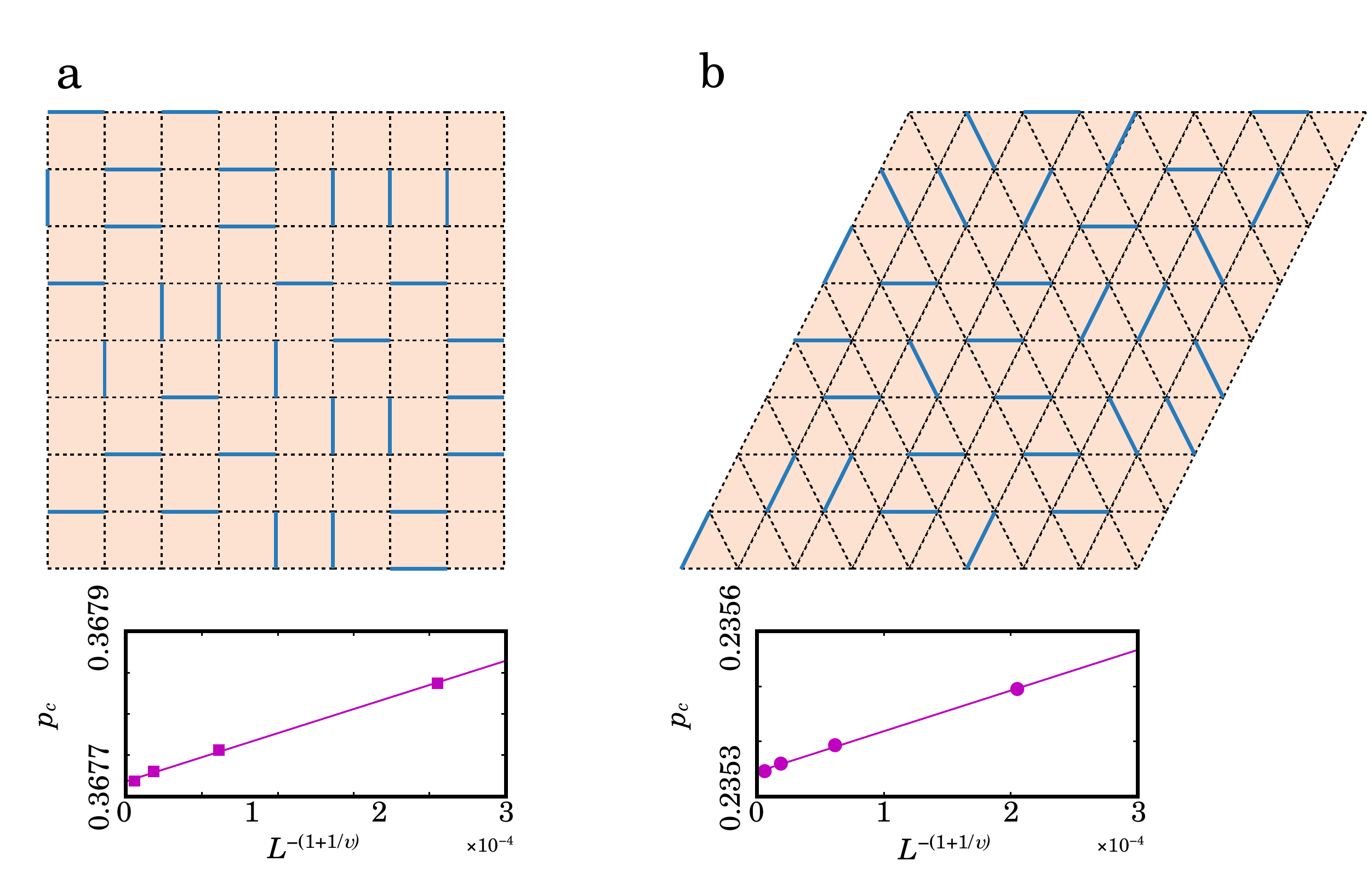}
    \caption{\label{fig:scheme}Schematic representation of random dimer coverings of (a) a square, and (b) a triangular lattice. The finite-size scaling of the computed percolation thresholds are given in the bottom.}
    \end{figure*}

    The algorithm that is used for generating random coverings of a lattice is based in creating closed paths of dimers, and shifting the permanently occupied bonds along that path. Starting with a given initial dimer covering, such as the parallel covering depicted in Fig.~\ref{fig:randomize}a, a sequence of shuffling moves are performed. In each move, a closed path $P$ is constructed as follows. First a site is randomly selected on the lattice, such as point 1 in Fig.~\ref{fig:randomize}b. Both the selected site and its dimer pair (i.e.,~point 2 in Fig.~\ref{fig:randomize}b) are added to $P$. The next step is to choose a second dimer by randomly selecting one of the $z-1$ unoccupied bonds ending in $2$, and adding it to $P$. Here $z$ is the coordination number of the original lattice. In Fig.~\ref{fig:randomize}b, for instance, 3 and 4 are added to $P$.  This procedure is continued until $P$ crosses itself at some point, e.g.~by returning to $3$ as in Fig.~\ref{fig:randomize}b. If the loop portion of $P$ contains an even number of sites {(which is alway the case for the square lattice but not the triangular lattice)}, the move is completed by flipping the bond identities along the closed portion of $P$, otherwise the move is rejected.    For instance, the move depicted in Fig.~\ref{fig:randomize}b gives rise to the covering depicted in Fig.~\ref{fig:randomize}c. Upon performing a sufficiently large number of moves, the covering becomes fully randomized, by losing correlation with the initial covering, and eventually transforms to a covering such as the ones depicted in Fig.~\ref{fig:scheme}. The random coverings used in $p_c$ calculations are gathered every $1,\!000$ steps, with each step consisting of $N_s$ moves. For every lattice size/type, we gather a minimum of 100 independent random dimer coverings.

\begin{figure*}
    \centering
    \includegraphics[width=\textwidth]{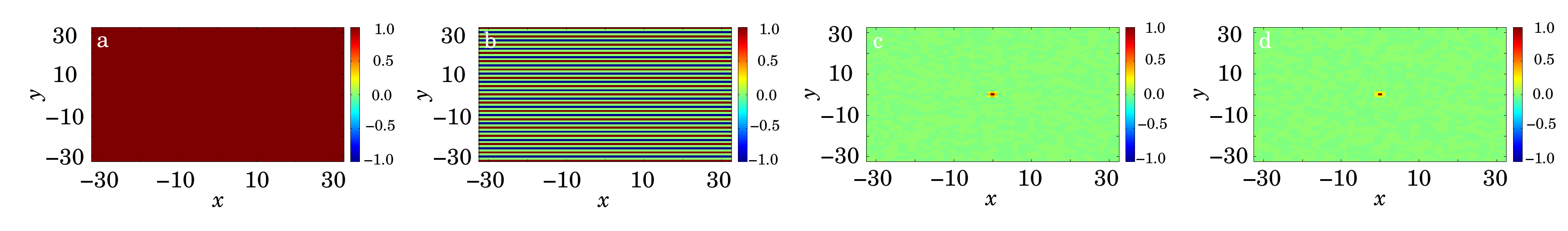}
    \caption{\label{fig:spatial}Spatial correlation on: (a) parallel and (b) staggered, (c) random square and (d) random triangular dimer-covered lattices. }
    \end{figure*}

    In order for our $p_c$ estimates to be reliable, it is necessary to establish the following two criteria: (a) In each random covering, no long-range spatial correlation should exist between the orientations of dimers.  (b) The successive coverings generated by the algorithm are to be uncorrelated.  In order to check for the first criterion, we compute a spatial orientational correlation function as follows. First, we assign to every site $i$ an orientation $o_i$. In the square lattice, the sites participating in horizontal and vertical dimers are assigned the values of $o_i=+1$ and $o_i=-1$, respectively. In the triangular lattice, values of $+1$, $-1$ and $0$ are assigned to horizontal, leftward and rightward dimers, respectively. Note that in a fully random covering, $\langle o_i\rangle=0$ since it must be equally likely for a site to belong to dimers of different orientations. The spatial orientational correlation function is then defined as:
    \begin{eqnarray}
    g_o(x,y) &=& \langle o_{i,j}o_{i+x,j+y}\rangle_{i,j}\label{eq:spatial}
    \end{eqnarray}
    with $i,j,x$ and $y$ the two-dimensional (integer) coordinates of the site on the lattice. In a fully-random covering of a lattice, $g_o(x,y)$ must decay exponentially along all arbitrary directions in the $(x,y)$ plane. Fig.~\ref{fig:spatial} depicts $g_o(x,y)$ for a representative random covering of a 128-by-128 square (Fig.~\ref{fig:spatial}c) and triangular (Fig.~\ref{fig:spatial}c) lattice. Note the rapid decay of $g_o(x,y)$ in all directions. $g_o(x,y)$ is virtually zero beyond the third nearest neighbor shell.  In contrast, the ordered lattices depicted in Figs.~\ref{fig:ordered:sq} and~\ref{fig:ordered:tr} tend to have non-decaying spatial correlations in all directions (Fig.~\ref{fig:spatial}a-b). This shows that the algorithm utilized here creates coverings that are random, and that lack no spatial correlations in the orientations of dimers. 
    
	To test (b), we compute the site orientational autocorrelation function given by:
    \begin{eqnarray}
    f_s(t) &=& \langle o_i(\tau)o_i(\tau+t)\rangle_{i,\tau}\label{eq:autocorrelation}
    \end{eqnarray}
    Here $t$ corresponds to the number of steps, with each step comprising of $N_s$ trial walks. We compute $f_s(t)$ for coverings of a $256\times256$ square and triangular lattice generated by the above mentioned algorithm. The decay in $f_s(t)$ is very rapid and $f_s(t)$ becomes virtually zero for $t>2$. This is much smaller than the $1,\!000$-step separation used for gathering random coverings. Therefore, the distinct random coverings used in this work are truly independent and lack no correlation. 
    
        \subsection{Percolation Thresholds}
All percolation thresholds are computed using the Newman-Ziff algorithm~\cite{NewmanZiffPRE2001}.  In each trial, a sequence of randomly selected unoccupied bonds are occupied one at a time. Upon the addition of each bond, the largest cluster of connected sites is updated. The procedure is continued until the largest cluster wraps around the lattice in either of the $x$ or $y$ directions. The percolation threshold is then estimated by $p_c=\langle N_m\rangle/N_{\text{all}}$ with $N_m$ being the number of bonds that have been occupied throughout the trial, and $N_{\text{all}}=N_b-(1/2)N_s$ the total number of (initially) unoccupied bonds. The averaging is taken over different trials. For every lattice, whether ordered or randomly covered, we perform $10^6$ trials, which gives a precision of three significant digits for $p_c$. The random covering percolation thresholds are obtained through an additional level of averaging over a minimum of 100 independent random coverings. This increases the precision in the computed $p_c^{\text{random}}$ to four significant digits.

The procedure outlined above gives $p_c$ on a finite lattice. However,  percolation transition is only precisely defined in an infinite system. There is a large body of work in the literature on how different estimates of $p_c$ on finite lattices scale with the characteristic length scale of the lattice. A detailed overview can be found in Ref.~\cite{ZiffNewman2002}. It follows from general scaling arguments that $|p_c(L)-p_{c,\infty}|\sim L^{-1/\nu}$ with $\nu=4/3$ for percolation in two dimensions. However, faster convergence can become possible depending on the  system shape, boundary conditions, the lattice, and the numerical method that is used for estimating $p_c$. In this work, we use both $|p_c(L)-p_{c,\infty}|\sim L^{-1/\nu}$ and $|p_c(L)-p_{c,\infty}|\sim L^{-(1+1/\nu)}$ for finite-size scaling.  The latter scaling applies for a system with special symmetry, such as  a square or rhombic shape, while the former applies for a general system.

    \subsubsection{High-precision  $p_c$'s from Epidemic Growth}
    
\begin{figure}
\centering
\includegraphics[width=.45\textwidth]{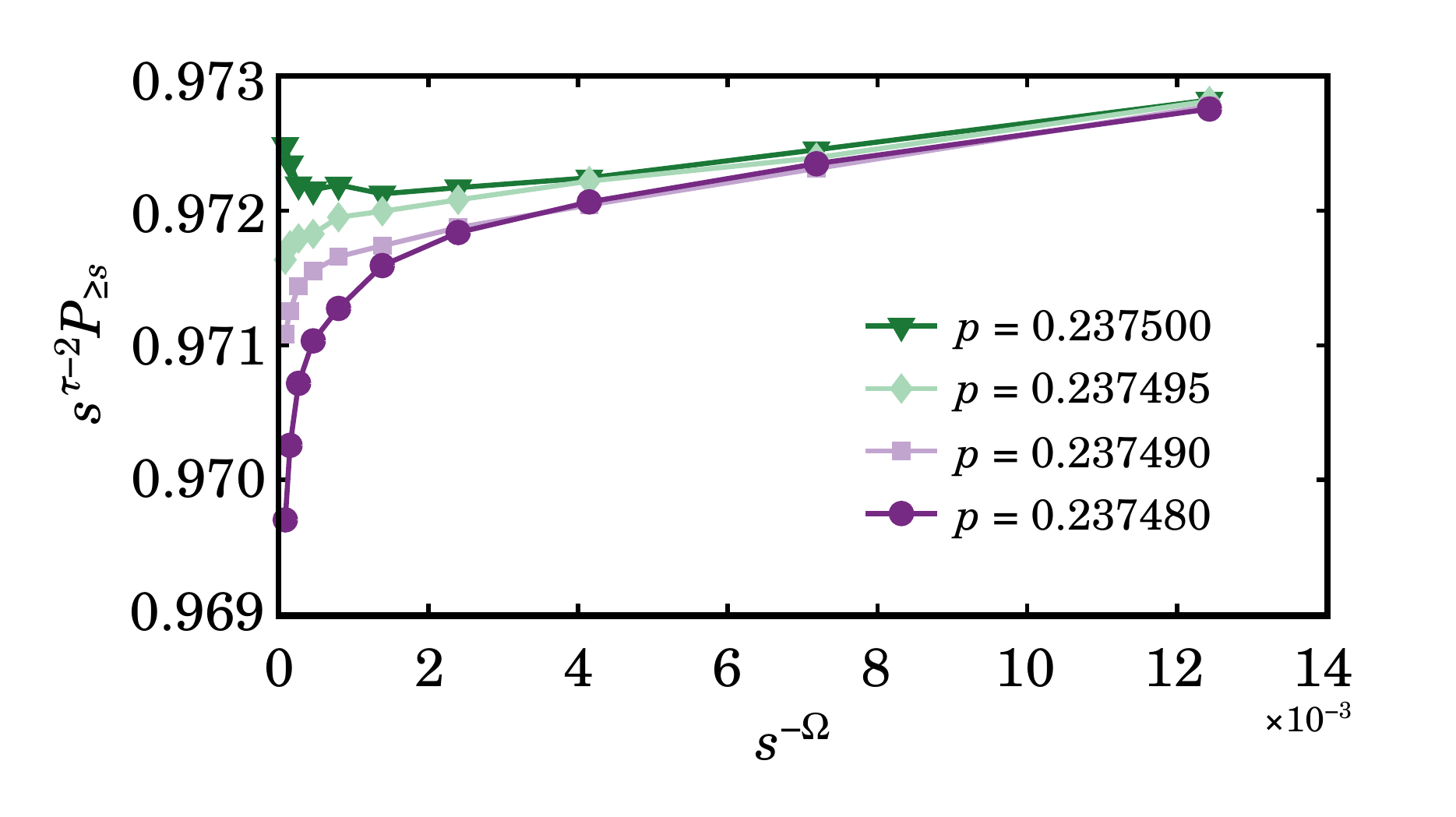}
\caption{\label{fig:lorenzziff} Application of the epidemic growth method for the staggered covering of the triangular lattice.}
\end{figure}

    To further increase the accuracy of the estimated percolation thresholds, we carry out simulations using the epidemic growth method of Lorenz and Ziff~\cite{LorenzZiffPRE1998}. In this method, a site $v$ is randomly chosen on the lattice, and a cluster is grown from $v$ via a Leath-type algorithm in which bonds are added at its periphery. Here, each  bond is occupied with probability $p$ and the process continues until there are no more growing bonds at the periphery, or when $2^{17}=131,\!072$ sites are already wetted by the growing cluster. Cluster size distribution data are accumulated in logarithmic bins $(2^n, 2^{n+1}-1)$, which are then used for estimating $P_{\ge{s}}$, the probability that the grown cluster is larger than $s=2^n$. We perform this on sufficiently large lattice ($8,\!192\times8,\!192$) so that the growing clusters never wrap around the periodic boundary. Therefore, our estimates of $P_{\ge{s}}$ are not affected by the finite size of the system.  According to the scaling theory, $P_{\ge s}$ scales as $s^{2-\tau}$ where $\tau = 187/91$ in two dimensions, and the finite-size scaling for small clusters is of the form
\begin{equation}
s^{\tau-2} P_{\ge s} \sim A + B s^{-\Omega}\label{eq:scaling}
\end{equation}
with $\Omega = 72/91$~\cite{AharonyAsikainenFractal2003, ZiffPREScaling2011}. This relationship holds for sufficiently large $s^{-\Omega}$.
When $p$ is away from $p_c$, there will be large deviations from the above behavior for large $s$. Fig.~\ref{fig:lorenzziff} depicts  $s^{\tau-2} P_{\ge s}$ vs.~$s^{-\Omega}$ for bond percolation on the staggered triangular lattice for values of $p$ close to where the large-$s$ behavior is linear. From this data we estimate $p_c=0.237497(2)$. This is consistent with the estimate of $p_c=0.237(1)$ obtained from the Newman-Ziff algorithm.  

For refining the error bars of the $p_c$'s estimated for random dimer coverings, we combine the above algorithm with partial rearrangements of dimers using a shuffling algorithm similar to the one explained in Section~\ref{section:shuffling}. For the square lattice, we continue the walks until the origin is revisited for the first time, and flip every bond along the path. For the triangular lattice, we flip bonds on a single intercepting loop, as long as the loop contains an even number of steps. Clusters are grown from random sites on the lattice. By having a sufficiently large lattice, we are able to grow an ensemble of uncorrelated clusters in different parts of the lattice without the need for completely rearranging the covering.   We also use a technique of labeling occupied sites by the index of the number of runs, with the sites with indices less that that considered vacant. This way, we can avoid clearing the lattice between successive clustering events.

For each covering, we start with a trial range of $p_{c,NZ}\pm2\delta p_{c,NZ}$, with $p_{c,NZ}$ the percolation threshold obtained from the Newmann-Ziff algorithm, and $\delta p_{c,NZ}$ the associated error bar. We choose several $p$ values in the interval, and by bisecting in between the $p$'s for which (\ref{eq:scaling}) becomes linear the earliest, we are able to narrow down $p_c$ for precisions as high as six significant digits. For each $p$, we carry out a minimum of $2\times10^7$ cluster growths. The longer simulations that are carried out for obtaining high-precision $p_c$'s can take up to several core-days on a computer.

    \section{Results and Discussion\label{section:results}}
    We compute the percolation thresholds of the ordered coverings of Figs.~\ref{fig:ordered:sq} and~\ref{fig:ordered:tr}, as well as the random coverings generated using the algorithms outlined above. The findings are summarized in Table~\ref{tab:summary}. The $p_c$'s computed for finite lattices scale linearly with either $L^{-1/\nu}$ or $L^{-1(1+1/\nu)}$, with $L$ the characteristic length scale of the lattice, i.e.,~the number of sites along each dimension. This allows us to accurately estimate the percolation threshold of an infinite lattice, as depicted in Figs.~\ref{fig:ordered:sq},~\ref{fig:ordered:tr} and \ref{fig:scheme}. The estimates of $p_c$ on all ordered coverings of the square lattice scale linearly with $L^{-1/\nu}$, while all random coverings (both on the square and the triangular lattice) as well as the two ordered coverings of the triangular lattice coverage more rapidly to $p_{c,\infty}$ with the scaling $L^{-(1+1/\nu)}$. For the three coverings with known percolation thresholds, i.e.,~the parallel and shifted coverings of the square lattice (Fig.~\ref{fig:ordered:sq}b), and the parallel covering of the triangular lattice (Fig.~\ref{fig:ordered:tr}a),  our numerical estimates are consistent with the known exact values of $p_c$, confirming the reliability of the $p_c$'s reported for the other ordered and random dimer coverings. 
    
    We use the $p_c$ values obtained from the finite-size scaling analysis as initial guesses for the epidemic growth method. As mentioned above, no finite-size correction is necessary for the refined $p_c$'s obtained from the epidemic growth method, due to the large lattice sizes considered therein. One can therefore test the accuracy of our initial finite-size scaling analysis by determining the exponent $c$ from:
    \begin{eqnarray}
    |p_{c,\infty}-p_c(L)| &=& aL^{-c} \label{eq:test_scaling}
    \end{eqnarray}
    Here $p_{c,\infty}$ and $p_c(L)$ are obtained from the epidemic growth algorithm, and the Newman-Ziff algorithm, respectively.  Table~\ref{tab:finite_exponent} summarizes the $c$ values obtained for different coverings in this work. These exponents are sufficiently close to the original $1/\nu$ and $1+1/\nu$ exponents used in our initial finite-size scaling analysis. Note that these exponents do not match completely with either of $1/\nu$ or $1+1/\nu$ because of numerics of considering finite number of $p_c(L)$'s in the fit, as well as the statistical uncertainty of the existing estimates. Also, one cannot rule out the possibility of a `cross-over' behavior, which will necessitate very large $L$'s for observing the true asymptotic behavior of $|p_{c,\infty}-p_c(L)|$.

    \subsection{Frustration of the Percolation Transition}
    As thoroughly discussed in Section~\ref{section:statement}, dimer covering of a lattice leads to a frustration of percolation transition in the original lattice, and the extent of such frustration can be quantified by computing $\kappa=p_{c,\text{eq}}/\tilde{p}_c$. Since $\tilde{p}_c$ is known exactly for the (uncovered) square and triangular lattices, $\kappa$ can be readily calculated from the computed $p_{c,\text{eq}}$ values.  The results are  given in Table~\ref{tab:summary}. The percolation transition is frustrated on dimer coverings of both lattices. However, the square lattice tends to be more adversely affected. The parallel and ordered coverings of the square lattice percolate at $p_{c,\text{eq}}^{\text{parallel}}=0.631966\cdots$ and $p_{c,\text{eq}}^{\text{staggered}}=0.626825(2)$, respectively, while the random covering percolates at a slightly lower $p_{c,\text{eq}}^{\text{random}}=0.617713(2)$. The least affected covering is the shifted covering, which percolates at $p_{c,\text{eq}}^{\text{shifted}}=0.597296\cdots$. Note that the frustration coefficient is above $1.19$ for all these coverings. Percolation frustration tends to be weaker in the triangular lattice, and $\kappa$ is always below 1.16. Also, the ordered and random covering percolation thresholds tend to be much closer, and are identical up to the second significant digit. This difference can be attributed to higher connectivity of the triangular lattice in which the permanently occupied dimer bonds constitute a smaller fraction of total bonds. Consequently, the emergence of an infinite cluster is expected to be less sensitive to the \emph{a priori} occupation of dimer bonds, as well their specific arrangement. It can henceforth be argued that percolation frustration will become even weaker in dimer coverings of lattices with larger coordination numbers. In two dimensions, such lattices can be constructed by adding multiple bonds between neighboring sites. Another possibility is to consider three-dimensional lattices, such body-centered cubic and face-centered cubic that have higher coordination numbers.

     \begin{table*}
    \centering
    \caption{\label{tab:summary}Percolation thresholds of dimer-covered lattices.}
    \begin{tabular}{lllll}
    \hline\hline
    Lattice~~~~~~~~~~~~& Dimer Arrangement~~~~~~ & $p_c$& $p_{c,\text{eq}}$ & $\kappa=p_{c,\text{eq}}/\widetilde{p}_c$\\
    \hline
    Square & Parallel & $0.381966\cdots$ & $0.631966\cdots
$ & $1.263932\cdots$\\
    Square & Shifted & $0.347296\cdots$ & $0.597296\cdots
$ & $1.1946\cdots$\\
    Square & Staggered & $0.376825(2)$ & $0.626825(2)$ & $1.25365(4)$\\
    Square & Random & $0.367713(2)$ & $0.617713(2)$ & 1.23542(4) \\
    Triangular & parallel & $0.237418\cdots$ & $0.404085\cdots$ & $1.163516\cdots$ \\
    Triangular & Staggered & $0.237497(2)$ & $0.404162(1)$ & 1.16374(2)\\
    Triangular & Random & $0.235340(1)$ & $0.402007(1)$ & 1.15753(2)\\
    \hline
    \hline
    \end{tabular}
    \end{table*}
    
    \begin{table}
    	\centering
		\caption{\label{tab:finite_exponent}Finite-size scaling exponent, $c$, for the coverings considered in this work.  }
		\begin{tabular}{llc}
		\hline\hline
		Lattice & Covering & Exponent \\
		\hline
		Square & Parallel & 0.90 \\
		Square & Shifted & 0.88\\
		Square & Staggered & 0.92\\
		Square & Random & 1.97\\
		Triangular & parallel & 1.32\\
		Triangular & Staggered & 1.38\\
		Triangular & Random & 1.72\\
		\hline
		\end{tabular}
    \end{table}
    
  	\subsection{Covering Order and Percolation}
   Another interesting question about percolation on dimer covered networks is the role of the covering order on percolation. Our findings reveal the difficulty of this task, as small changes in the covering can translate into significant changes in the percolation threshold. For instance, the parallel and shifted coverings of the square lattice are almost identical, except for a shift in the dimer positions every other row. However, they tend to have widely different percolation thresholds. Indeed, their $p_c$'s of $0.381966\cdots$ and $0.347296\cdots$ tend to fall on the opposite ends of the random covering percolation threshold $p_c^{\text{random}}=0.367713(2)$. This granularity underscores the difficulty of identifying an order parameter for characterizing the role of covering order on percolation, as simple order parameters such as $\langle o_i\rangle$ are inadequate for that purpose. The origin of this granularity is interesting on its own. The shifted covering seems to be a percolation `sweet spot'  on the space of all possible coverings of the square lattice. This can be understood by inspecting the topologically equivalent networks of ordered and random coverings of the square lattice. All these coverings have the same coordination number of $z=6$. However, the parallel (Fig.~\ref{fig:ordered:sq}a) and the staggered (Fig.~\ref{fig:ordered:sq}b) covering, as well as the random coverings tend to have a large number of double bonds in their topologically equivalent network, while the shifted covering is exclusively comprised of single bonds. Therefore, a single dimer on the shifted covering can be connected to a \emph{larger} number of distinct dimers than a similar dimer on the parallel, staggered and random coverings. In other words, the stochastically occupied connections of the shifted network are distributed more efficiently as to allow the largest number of site-to-site connections at a particular occupation probability. It can therefore be conjectured that among the two-dimensional six-coordinated networks, triangular networks (whether regular or irregular) will have the smallest percolation thresholds.  Here, it is necessary to mention a related conjecture due to Wierman that states that the regular triangular lattice has the highest bond percolation threshold among all triangulated networks~\cite{WiermanJPhysA2002}. 
   
   Unlike the square lattice, the difference between ordered and random coverings of the triangular lattice is very small, and the corresponding percolation thresholds differ only at their third significant digit. Also, the two ordered coverings considered here have a higher $p_c$ than the random covering. In other words, no percolation sweet spots are found among the dimer coverings of the triangular lattice. Indeed, no such ordered covering might exist as no planar two-dimensional lattice is known to have a coordination number of $z=10$, which is the coordination number of the topological networks of the coverings of the triangular lattice.  
   
   	\subsection{Comparison with Other Random Networks}
	
	Before comparing the percolation characteristics of random dimer coverings with other random networks that are defined using the notion of a norm (i.e.,~distance between points in an Euclidean space),  it is necessary to make a very important distinction. In the norm-based random network such as the relative-neighborhood graph or the Voronoi tessellation, the coordination number of each site is a correlated random variable, although uniformity arguments can be made about the average coordination numbers based on the stochastic distribution of the generating vertices. In the covering-based random networks, however, each site can take a finite number of coordination numbers that depend on the coordination numbers of the original lattice. To be precise, consider a dimer bond that connects two sites with  coordination numbers $z_1$ and $z_2$. The corresponding unified dimer site on the associated topologically equivalent network will  have a coordination number of $z_{12}=z_1+z_2-2$. This confines the number of allowable coordination numbers to a set of known $z_{ij}$'s. In the square and the triangular lattice, all sites have the same $z$, so all sites on the covering-based random network will be $(2z-2)$-coordinated. Therefore, dimer covering can be viewed as a systematic way of generating random networks with precisely determined coordination numbers.

    Considering the large coordination numbers of the covering-based random networks, it is not straightforward to compare their percolation characteristics with those of the known two-dimensional norm-based random networks, such as the relative-neighborhood graph,  Voronoi tessellation and the Gabriel graph that have $\bar{z}=2.5576$~\cite{MelchertPRE2013}, $\bar{z}=3$~\cite{BeckerPRE2009} and $\bar{z}=4$~\cite{BertinAdcApplProv2002}, respectively. Not surprisingly, these low-coordination graphs tend to have higher percolation thresholds, descending in the order of  increase in $\bar{z}$. The only comparison can be made between the Delaunay triangulation ($p_c=0.333069$)~\cite{BeckerPRE2009} and the random covering of the square lattice $p_c=0.367713(2)$ that are both six-coordinated. Indeed, a random covering of the square lattice will have a considerable number of double bonds, which, as explained above, are likely to lead to higher percolation thresholds.  In the case of the ten-coordinated triangular lattice, no two-dimensional random network with a similar coordination number has been studied.

    
    \section{Conclusion\label{section:conclusion}}
    In this work, we report the first computational investigation of bond percolation on ordered and random dimer coverings of the square and the triangular lattice. For generating random coverings, we propose a novel shuffling algorithm that is conceptually similar to an algorithm used for generating proton-disordered arrangements of ice~\cite{StillingerJCP1972}. The percolation transition is strongly frustrated in both lattices as a result of dimer covering, with the frustration being stronger in the low-coordination square lattice. We also observe that there is no clear correlation between the existence of long-range order in a covering and its percolation characteristics, as closely-related ordered coverings can have $p_c$'s on opposite ends of the random covering percolation threshold. The ordered and random coverings of the triangular lattice considered in this work have very close percolation thresholds suggesting that the microscopic detail of the covering is less likely to affect the percolation characteristics in highly-coordinated lattices and networks. We also compare random coverings with norm-based random lattices such the Voronoi tessellation and Delaunay triangulation both in terms of connectivity and the percolation threshold. 
    
This work paves the way for studying a wide range of other interesting questions about percolation on dimer-covered lattices. Obviously, it is interesting to study bond percolation on dimer coverings of other two- and three-dimensional lattices. One interesting question there is the sensitivity of percolation frustration to the coordination number, dimensionality and the microstructure of the corresponding lattice. As suggested here, and as expected intuitively, frustration is expected to be smaller in lattices with higher coordination numbers. However, the effect of dimensionality and microstructure (i.e.,~the particular local connectivities in the lattice) is less straightforward to predict. Also, the idea of frustration can be  used for understanding and describing the behavior of permanently connected systems, such as entanglement in polymeric systems.  

Another question of interest is to study dimer coverings in which dimer bonds are permanently vacated. Such a model can be used for generating random coverings with low coordination numbers. For instance, the vacant covering of the square and the triangular lattice lattice will have a coordination number of $z=3$ and $z=5$, respectively. Finally, one can consider the problem of $n$-mer covering, which is an extension of the dimer covering problem.

    \acknowledgments 
    
    These calculations were performed on the Terascale Infrastructure for Groundbreaking Research in Engineering and Science (TIGRESS) at Princeton University. This research was also supported in part through computational resources and services provided by Advanced Research Computing at the University of
Michigan, Ann Arbor.
    
    \bibliographystyle{apsrev}
    \bibliography{References}

    \end{document}